\documentclass[aps,prd,twocolumn,groupedaddress,nofootinbib]{V63/aastex63}
\usepackage{graphicx,color,amsmath}
\usepackage{slashed}
\usepackage{booktabs} 
\usepackage{braket}
\usepackage{hyperref}
\usepackage{mathtools} 

\usepackage{tikz}
\usepackage{tkz-euclide}
\usetikzlibrary{decorations.text}
\usetikzlibrary{decorations.pathmorphing}	
\usetikzlibrary{fadings}
\tikzset{
    v/.style={decorate, decoration={snake, segment length=3mm, amplitude=0.75mm}, draw},
    f/.style={draw,decoration={markings,mark=at position #1 with {\arrow[very thick]{latex}}},postaction={decorate},node contents=#1},
    f/.default=.6,
    fb/.style={draw,decoration={markings,mark=at position #1 with {\arrowreversed[very thick]{latex}}},postaction={decorate},node contents=#1},
    fb/.default=.4,
    fnar/.style={draw},
    g/.style={decorate, draw,  decoration={coil,amplitude=3pt, segment length=3.5pt}},
    s/.style={dashed,draw, postaction={decorate},
        decoration={markings,mark=at position .55 with {\arrow[very thick]{latex}}}},
    sb/.style={dashed,draw, postaction={decorate},
        decoration={markings,mark=at position .55 with {\arrowreversed[draw=black,very thick]{latex}}}},
    snar/.style={dashed,draw,line width =1.25pt},
}

\newcommand{\beq}{\begin{equation}}
\newcommand{\eeq}{\end{equation}}
\newcommand{\ba}{\begin{array}}
\newcommand{\ea}{\end{array}}
\newcommand{\bea}{\begin{eqnarray}}
\newcommand{\eea}{\end{eqnarray}}

\newcommand{\bal}{\begin{align}}
\newcommand{\eal}{\end{align}}
\newcommand{\bi}{\begin{itemize}}
\newcommand{\ei}{\end{itemize}}
\newcommand{\ben}{\begin{enumerate}}
\newcommand{\een}{\end{enumerate}}
\newcommand{\bc}{\begin{center}}
\newcommand{\ec}{\end{center}}
\newcommand{\bt}{\begin{table}}
\newcommand{\et}{\end{table}}
\newcommand{\btb}{\begin{tabular}}
\newcommand{\etb}{\end{tabular}}

\newcommand{\msun}{\ensuremath{{\rm M}_\odot}}
\newcommand{\rsun}{\ensuremath{{\rm R}_\odot}}
\newcommand{\lsun}{\ensuremath{{\rm L}_\odot}}

\providecommand{\keywords}[1]
{
  \textbf{Keywords---} #1
}

\begin{document}

\title{A Measurement of Stellar Surface Gravity Hidden in Radial Velocity Differences of Co-moving Stars}


\author{Matthew Moschella}
\affiliation{Department of Physics, Princeton University, Princeton, NJ 08544, USA}

\author{Oren Slone}
\affiliation{Department of Physics, Princeton University, Princeton, NJ 08544, USA}
\affiliation{Center for Cosmology and Particle Physics, Department of Physics, New York University, New York, NY 10003, USA}

\author{Jeff A. Dror}
\affiliation{Department of Physics and Santa Cruz Institute for Particle Physics, University of California, Santa Cruz, CA 95064, USA}
\affiliation{Theory Group, Lawrence Berkeley National Laboratory, Berkeley, CA 94720, USA}
\affiliation{Berkeley Center for Theoretical Physics, University of California, Berkeley, CA 94720, USA}

\author{Matteo Cantiello}
\affiliation{Center for Computational Astrophysics, Flatiron Institute, 162 5th Avenue, New York, NY 10010, USA}
\affiliation{Department of Astrophysical Sciences, Princeton University, Princeton, NJ 08544, USA}

\author{Hagai B. Perets}
\affiliation{Faculty of Physics, Technion – Israel Institute of Technology, Haifa, 3200003, Israel}

\nocollaboration{5}

\correspondingauthor{Oren Slone}
\email{oslone@princeton.edu}

\begin{abstract}
The gravitational redshift induced by stellar surface gravity is notoriously difficult to measure for non-degenerate stars, since its amplitude is small in comparison with the typical Doppler shift induced by stellar radial velocity. In this study, we make use of the large observational data set of the {\em Gaia} mission to achieve a significant reduction of noise caused by these random stellar motions. By measuring the differences in velocities between the components of pairs of co-moving stars and wide binaries, we are able to statistically measure the combined effects of gravitational redshift and convective blueshifting of spectral lines, and nullify the effect of the peculiar motions of the stars. For the subset of stars considered in this study, we find a positive correlation between the observed differences in {\em Gaia} radial velocities and the differences in surface gravity and convective blueshift inferred from effective temperature and luminosity measurements. The results rule out a null signal at the $5\sigma$ level for our full data-set. Additionally, we study the sub-dominant effects of binary motion, and possible systematic errors in radial velocity measurements within {\em Gaia}. Results from the technique presented in this study are expected to improve significantly with data from the next {\em Gaia} data release. Such improvements could be used to constrain  the mass-luminosity relation and stellar models which predict the magnitude of convective blueshift.\\\\
\end{abstract}



\section{Introduction}

The advent of the {\em Gaia} space telescope has given rise to a new era of precision astrometry with a current catalog that includes over one billion stars, of which several millions have radial velocity (RV) measurements. Such a large data set offers new opportunities to statistically measure stellar properties. In this study, we use RV measurements from {\em Gaia} to measure the gravitational redshift (GR) due to stellar surface gravity (SG), as well as the subdominant effect of convective outflows and downflows at the stellar surface.

Understanding these effects is important since they give rise to systematic noise in RV measurements, and in particular raise difficulties in RV detection of exoplanets \citep[see e.g.][for a review]{2018haex.bookE...4W}. Additionally, measurement of convective effects could shed light on physical processes occurring within stars and constrain current and future modeling of these processes.

\begin{figure*}[t]
  \centering
  \includegraphics[width=.49\textwidth]{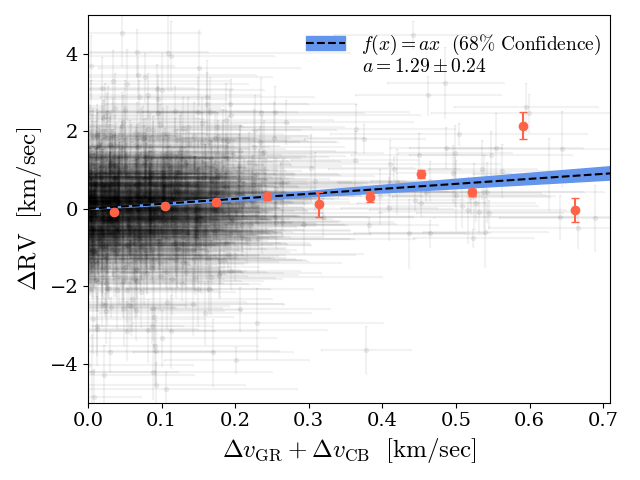}
  \includegraphics[width=.49\textwidth]{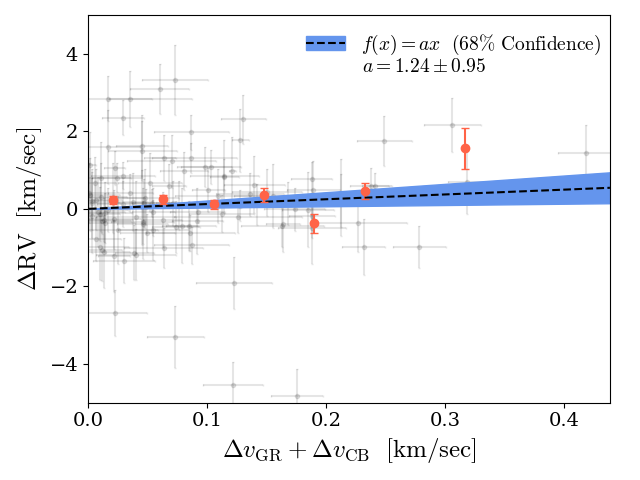}
  \caption{The observed correlation between differences in radial velocity measurements between co-moving stellar pairs and the expected difference due to gravitational redshift and the smaller effect of convective blueshift. {\bf Left panel:} Stars that passed the selection criteria but do not have precise spectroscopic mass measurements (primary data set). Since mass measurements are imprecise for this data set, convective blueshift modeling is challenging. {\bf Right panel:} The subset of co-moving pairs which have precise mass measurements (secondary data set). For each panel, a model which perfectly fits the data would correspond to a linear fit of the form $f(x)=x$, with a slope of unity. The observed best fit linear curves are given as insets in each panel. The result for the primary (secondary) data set is consistent with a non zero slope at the $\sim5\sigma$ ($\sim1.5\sigma$) level. The large error for the secondary data set is mainly due to low statistics.
    \label{fig:money_plot}
    }
\end{figure*}

The {\em Gaia} mission performs RV measurements by measuring the Doppler shift of a few common absorption lines using the Radial Velocity Spectrometer (RVS). Although the frequency shift observed by the RVS is reported as a relative velocity between the emitting star and the {\em Gaia} satellite, there are many additional effects that can contribute \citep[see \textit{e.g.},][for an overview]{Lindegren:2003wg}. In particular, the SG of the emitting star results in a redshift of line frequencies measured by the RVS, biasing RV measurements to be slightly more positive than their true values. While the true velocities are typically on the order of the galactic virial velocity, $v_{\rm gal}\sim 220$ km/sec, typical values from GR are,
\beq
v _{ {\rm GR}} \approx 0.6 \left(\frac{M}{\msun}\right)\left(\frac{\rsun}{R}\right) \frac{{\rm km}}{{\rm sec}} \label{eq:v_SG_val}\,.
\eeq
Thus, GR in non-degenerate stars is usually a small contribution to the RV.

An additional effect on RV measurements arises from convective motion on the stellar surface. Hot and luminous outflows typically result in a net blueshift of emitted photons and measured RVs appear, on average, smaller (more negative) than their true values. This effect is known as convective blueshift (CB). The size of the effect depends on stellar type and can induce effective RV measurements of order, $v _{ {\rm CB}} \sim \mathcal{O}(0.3\text{ km/sec})$, and hence is generally subdominant to GR.

Although existing spectrometers have the necessary precision to observe GR, this measurement is challenging since the true velocity of any particular star is unknown, contributing a statistical uncertainty on the order of $v_{\rm gal}$. This difficulty has been overcome in only a few types of systems. Original measurements of GR were performed on white dwarfs in binary systems, most notably the nearby Sirius B~\citep{1971ApJ...169..563G}. These can have GRs as large as ${\cal O} (100~{\rm km}/{\rm sec})$~\citep{1967ApJ...149..283G}, making detection relatively straightforward. Additionally, there have been long-standing efforts to extract the GR of the Sun, which is expected to be $636.5\, {\rm m}/{\rm sec}$~\citep{Lindegren:2003wg}. This is challenged by a CB of comparable size, which itself depends sensitively on the spectral line and its angular position. A recent clean measurement and some discussion of its history can be found in~\cite{Hernandez:2020wdy}.

The difficulty of disentangling GR from physical motion in stars can be reduced with statistics of large data sets and an appropriate choice of stellar system. A previous attempt to statistically measure GR in ordinary stars was performed using the M67 open clusters in~\cite{2011A&A...526A.127P}. However, that study was unable to extract a signal above the background, which the authors hypothesized as due to unsubtracted CB effects. More recently, ~\cite{Le_o_2018} compared spectroscopic and astrometric RVs for stars in the Hyades open cluster. That study was able to measure the combined effects of GR and CB, however their data sample was small and included only two giants. Finally~\cite{Dai:2018lwv}  attempted a measurement of stellar GR by averaging {\em Gaia} RV measurements over stars of different types. Their results show some evidence for a combination of GR and CB and they did not attempt to distinguish between or characterize the different contributions.

In this study, we search for GR in co-moving pairs of stars, thousands of which have been identified in \textit{Gaia} DR2. If these co-moving pairs are wide binaries, their orbital velocities should be small, providing a clean data set with which to study small contributions to RVs. In particular, the difference of the RVs of two stars in a wide binary can be dominated by the difference in $v_{\rm GR}$, especially for pairs of stars with sizeable differences in mass and/or radius.
This opens the possibility of using wide binaries to directly probe GR and indirectly gain information regarding stellar structure and dynamics.

The main results of this study are presented in Fig.~\ref{fig:money_plot}, which shows the correlation between the differences of the observed RVs of stars in co-moving pairs and the expectation from GR and CB. The left panel corresponds to a data set with higher statistics but also high uncertainty on stellar mass, while the right panel corresponds to a smaller sample with high quality mass measurements. A positive correlation is observed, corresponding to an agreement between model and data within measurement errors, with a non-zero slope at the $\sim1.5\sigma - 5\sigma$ level.

The study is structured as follows. First, we present details of the data selection used for this study. We follow with a discussion of GR- and CB-induced changes in measured RVs. Next, we present our analysis technique and discuss the dispersion observed in the data, concluding with a discussion of our main results, possible issues, and future prospects.

\section{Data Selection}

Comoving pairs in physical (3-dimensional) velocity were first compiled using {\em Gaia} DR1 by~\cite{2017AJ....153..257O}. Such stars are typically wide binaries with separation distances $\gtrsim 25~{\rm AU}$~\cite{2000A&A...356..529A}, and are distinct from typical binaries in that they are likely to have been born at the same time and share a similar chemical composition. This has led to considerable interest in compiling lists of wide binaries with 3741 pairs identified in DR2~\cite{2019AJ....157...78J}, and statistical analyses performed by \cite{2018arXiv181013270Z}. Additionally,~\cite{Hartman_2020} created the SUPERWIDE catalog of wide binaries. This catalog is selected from {\em Gaia} DR2 and the SUPERBLINK high proper motion catalog~\citep{2005AJ....130.1247L,2011AJ....142..138L}. Importantly, selection of these wide binary pairs is completely independent of RV measurements.

We analyze the subset of SUPERWIDE pairs that also have RV measurements in \textit{Gaia} DR2. Additionally, we require that the difference in the RVs of each star in a pair, $\Delta \mathrm{RV}$, has a small measurement error, $\sqrt{\sigma^2_{\mathrm{RV},1} + \sigma^2_{\mathrm{RV},2}}<1\ \mathrm{km/sec}$, and we remove outlier pairs with $\left|\Delta\mathrm{RV}\right|>5\ \mathrm{km/sec}$. To reduce contamination from the orbital velocities of wide binaries, we require that the transverse separation of the stars in a pair, $d$, be greater than $10^{-3}\ \mathrm{pc}$. Finally, we require that all stars have estimated values for effective temperature, $T_{\rm eff}$, luminosity, $L$, and radius, $R$, from {\em Gaia} DR2. A total of 1,682 pairs pass these selection criteria.

\begin{figure}
  \centering
  \includegraphics[width=.49\textwidth]{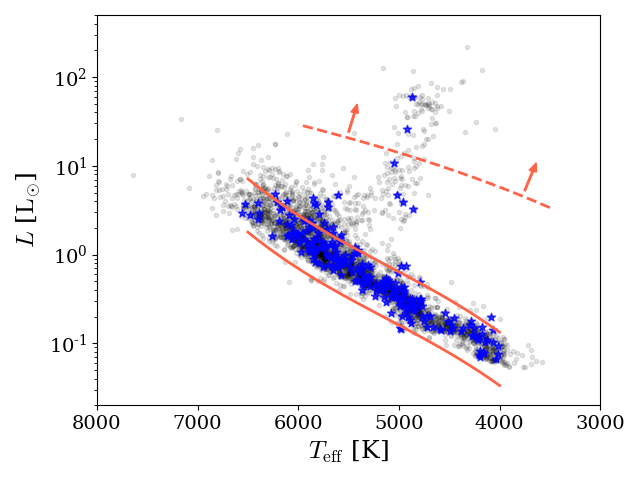}
  \caption{The Hertzsprung-Russell Diagram for the data sets used in this study. Grey points correspond to all SUPERWIDE stars that passed the basic quality cuts. Of these, stars in the region between the solid red curves are classified as main sequence and stars above the dashed red curve are classified as giants. Pairs for which either star is outside these regions are removed from the primary data set. Blue points correspond to stars in pairs for which both have spectroscopically determined mass estimates (secondary data set).
  \label{fig:HRD}
    }
\end{figure}

Since we are interested in estimating the masses and radii of stars in our data set with high accuracy, it is beneficial to categorize stars as either main sequence (MS) stars or as giants. The selection criteria used in this study are marked in red in Fig.~\ref{fig:HRD}. MS stars are defined to lie within the solid curves.\footnote{These curves are a multiplicative factor of $2$ from an $L(T_{\rm eff})$ curve which corresponds to the luminosity at which the highest number density 
is obtained across all stars in \textit{Gaia} DR2 that pass the selection criteria suggested in \cite{Babusiaux2018}. The curve is restricted to $4000\ \mathrm{K}<T_{\rm eff}<6500\ \mathrm{K}$ and $L>10^{-1.2}\lsun$ to ensure that the mass-luminosity relation from \cite{Malkov07} is valid. These criteria also assist in excluding wide binaries that are in fact triple systems containing an unresolved inner binary, 
which otherwise might contaminate the sample.} Giants are defined to lie above the dashed curve, which corresponds to $R>5\ \rsun$. A total of 1,135 pairs pass these additional selection criteria in our primary data set, of which 1,080 contain two MS stars, 53 contain one MS star and one giant, and 2 contain two giant stars.

We estimate the masses of MS stars using the mass-luminosity relation,
\begin{equation}
    \log_{}\left[\frac{M}{\msun}\right] \approx \alpha_0 + \alpha_1\log_{}\left[\frac{L}{\lsun}\right] + \alpha_2 \log^2_{}\left[\frac{L}{\lsun}\right],
\label{eq:M_L_relation}
\end{equation}
where $\alpha_0=0.00834$, $\alpha_1=0.213$, and $\alpha_2=0.0107$ \citep{Malkov07}.
However, we note that there is considerable uncertainty on the precise form of this relation. We conservatively assume an uncertainty of $0.1\ \msun$ in the masses estimated using Eq.~\eqref{eq:M_L_relation}.

For giant stars, the stellar mass can  in principle be inferred directly from photometric observations via asteroseismology. However, this requires long photometric observation times and cannot be done with \textit{Gaia} photometry. In order to preserve a sizeable statistical sample of giants, we uniformly set the masses of all selected giants to the average asteroseismologically determined mass from the \textit{Kepler} mission, $1.3 \msun$ \citep{Yu_2018}. We estimate the uncertainty on the fixed giant mass to be $0.5 \msun$. This \textit{ad hoc} estimation of the giant masses has a negligible effect on the GR signal in a MS - giant comoving pair, since the radius selection cut of $R>5~\rsun$ ensures that the GR contribution from giants is $\lesssim 20\%$ of that from MS stars.

In addition to the primary data set discussed above, we also identify a distinct set of comoving pairs for which both stars have high quality spectroscopic mass and radius measurements from~\cite{2018MNRAS.481.4093S}. Stars from this set of 114 pairs are marked in blue in Fig.~\ref{fig:HRD}. In principle, some of these star's RVs could also be measured directly from their spectra instead of the Gaia RVS, however we do not consider this possibility in the current study since the small statistics of the secondary data set is extremely limiting anyway.

\begin{figure}
  \centering
  \includegraphics[width=.45\textwidth]{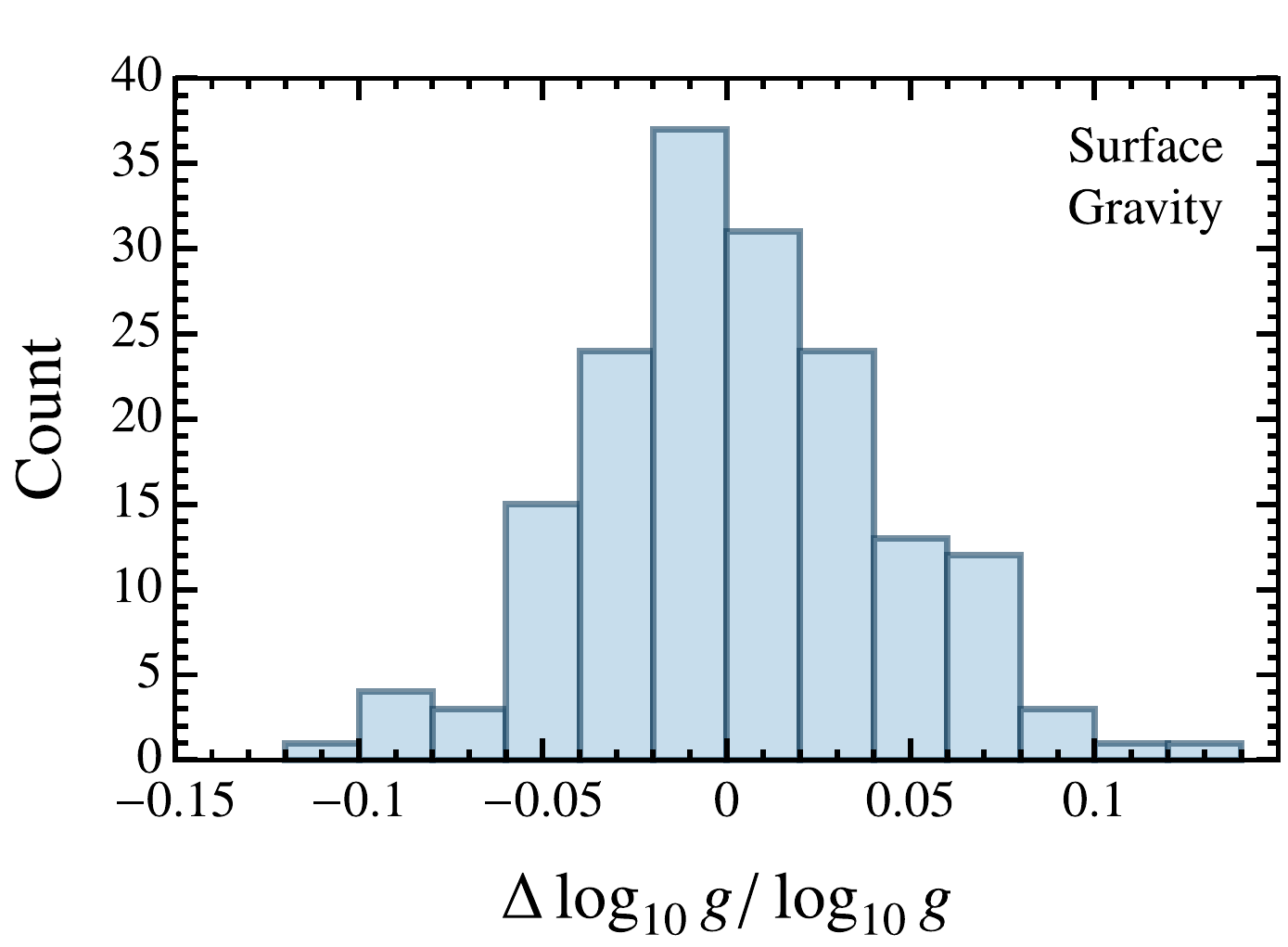}
  \caption{Histogram of the differences in surface gravity, $\log_{10}$g, as reported by Gaia (primary data set) and by~~\cite{2018MNRAS.481.4093S} (secondary data set) for stars which are in both data sets. The difference in normalized to the spectroscopic value from the secondary data set. We find that the differences in measurements of surface gravity between this data set and our primary data set are of order $\mathcal{O}(10\%)$ or less.
  \label{fig:logg}
    }
\end{figure}

To test consistency between the data sets, Fig.~\ref{fig:logg} shows a comparison of values of SG, $\log_{10}$g, between the data sets, as reported by Gaia and~\cite{2018MNRAS.481.4093S}. Specifically, for all star that are in both data sets, we plot a histogram of the differences in SG normalized to the spectroscopic value. We find that the differences in measurements of SG between the primary and secondary data set are of order $\mathcal{O}(10\%)$ or less. Tables of additional properties of stars in each pair of both data sets is available in the online Supplemental Material of this study through the journal webpage.

With these two data sets at hand, we proceed to estimate the effects of differences in GR and CB between stars in each pair.\\

\section{Gravitational Redshift}
\label{sec:GR}
The difference in measured RV between two stars (labelled 1 and 2), induced by GR alone, is given by
\begin{equation}
\Delta v_{\rm GR} = G \left(\frac{M_1}{R_1} - \frac{M_2}{R_2}\right).
\label{eq:deltavGR}
\end{equation}
Since giants have considerably larger radii, they tend to contribute negligibly to GR. Additionally, the GR of two MS stars tends to partially cancel. Thus, the largest signal comes from MS - giant pairs.

The masses and radii in Eq.~\eqref{eq:deltavGR} are estimated as described above (depending on the data set being used). For each co-moving pair, a distinct value of $\Delta v_{\rm GR}$ can be calculated and compared to the measured $\Delta$RV of the pair. This comparison is complicated by the presence of sub-dominant effects such as CB.

\section{Convective Blueshift}\label{sec:cb}
The surface of a star is characterized by the presence of convective cells where hot gas rises outwards, and cooler gas sinks through intergranular lanes. Since the hotter rising gas is brighter than the cooler sinking gas, a net blueshift is produced in the disk-integrated light emitted by the star. This phenomenon is usually referred to as CB \citep[e.g.][]{Beckers:1978,Dravins:1982}.

Since CB depends on  complex details of turbulent convection, it is notoriously difficult to predict. This presents a major challenge in, for example, exoplanet searches looking for modulations in spectral lines due to planetary motion around distant stars \citep[see e.g.][for a review]{2018haex.bookE...4W}. Attempts to estimate the amplitude of this effect for different stellar types rely on hydrodynamic simulations of surface convection. Such calculations were performed  in~\cite{2013A&A...550A.103A} for typical stellar types in the {\em Gaia} catalog, finding amplitudes of CBs in the range $v_{\rm CB} \approx 0.2-0.5~ {\rm km}/{\rm sec} $, depending on the mass and temperature of the star. While magneto-hydrodynamic calculations might provide a more complete picture, recent measurements show that magnetic activity is only mildly correlated with CB~\citep{refId0,2017A&A...597A..52M}. 

To parametrize the effect of CB we define the difference between stars 1 and 2 as $\Delta v_{\rm CB} \equiv v_{{\rm CB},1}-v_{{\rm CB},2}$ (note that $v_{{\rm CB},i}$ are typically negative) and apply the theoretical fitting formula provided in \cite{2013A&A...550A.103A}. This fit requires estimates of both metallicity and SG, which are not reported in {\em Gaia} spectroscopic data. For our primary data set, we estimate metallicity using the \textit{Gaia} RV template metallicity, and SG using the mass and radius estimates described above. For our secondary data set, we use metallicity and SG measurements reported in \cite{2018MNRAS.481.4093S}. Although the dependence of CB on metallicity is relatively weak, $\Delta v_{\rm CB}$ is highly sensitive to SG. Therefore, the use of Eq.~\eqref{eq:M_L_relation} may introduce sizeable errors in $\Delta v_{\rm CB}$ for the primary data set.

\section{Analysis}
For each pair in our data sets we calculate $\Delta v_{\rm GR}$ according to Eq.~\eqref{eq:deltavGR} and $\Delta v_{\rm CB}$ using the fitting formula of~\cite{2013A&A...550A.103A}. For stars in our primary data set, we estimate stellar masses and radii as described above; for our secondary data set we use the mass and radius estimates provided in \cite{2018MNRAS.481.4093S}.

\begin{figure*}[t]
  \centering
  \includegraphics[width=0.99\textwidth]{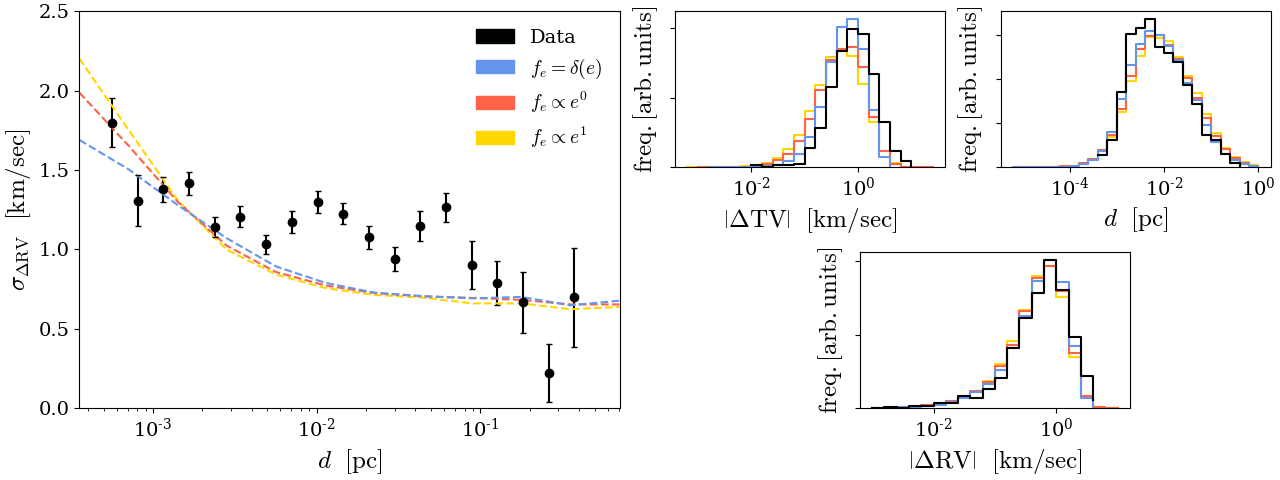}
  \caption{{\bf Left panel:} The dispersion in $\Delta$RV as a function of projected separation for the data used in this study compared with Monte Carlo mock data of wide binaries. {\bf Right panels:} Distributions of $\lvert\Delta {\rm RV}\rvert$, $\lvert\Delta{\rm TV}\rvert$ and $d$. For the mock data, the distribution of eccentricities is modeled either as a power law of the form $f_e \propto e^\eta$ or as circular orbits for all binaries. Differences resulting from the choice of eccentricity distribution are negligible and the consistency of the real data and mock data distributions in the right panels validate the Monte Carlo procedure. We find that the dispersion in the data is likely dominated by binary motion at low $d$, measurement errors at large $d$, and some unknown source at intermediate $d$.
   \label{fig:delta_rv_hist}
    }
\end{figure*}

In the absence of background effects, one expects $\Delta$RV to be directly correlated with $\Delta v_{\rm GR}+\Delta v_{\rm CB}$. This relationship is shown in Fig.~\ref{fig:money_plot}. The left panel corresponds to the primary data set, for which the lack of high quality mass estimates introduces large errors in $\Delta v_{\rm GR}$ and $\Delta v_{\rm CB}$. The right panel corresponds to the secondary data set for which $\Delta v_{\rm GR} + \Delta v_{\rm CB}$ is more precise but statistics are much lower. Grey points represent the pairs in each data set along with their corresponding uncertainties. The ordering of stars in each pair is chosen such that the value on the horizontal axis is positive. 
Orange points correspond to the average value of $\Delta$RV in bins of $\Delta v_{\rm GR} + \Delta v_{\rm CB}$, and are shown for illustrative purposes only.

In each case we test how well the model (horizontal axis, denoted $x$) fits the data (vertical axis, denoted $y$). If all background effects averaged to zero, one would expect a linear fit of the form $y=f(x)$ with $f(x)=a x$ and $a=1$. 
We find that this linear model is a poor fit to both data sets, mainly due to a dispersion of the measured $\Delta$RV values which is larger than that expected from measurement errors alone. That is, we find that there is an intrinsic source of dispersion in our $\Delta$RV measurements. We discuss possible sources of this dispersion below and account for it by introducing an additional model parameter, $\sigma_0$. We then analyze the data using the Gaussian likelihood,
\begin{equation}
\mathcal{L}(a,\sigma_{0}|\boldsymbol{x},\boldsymbol{y}) \propto \frac{1}{\sqrt{\prod_i \tilde{\sigma}_i^2}}\exp\left[-\frac{1}{2}\sum_i \frac{(y_i - a x_i)^2}{\tilde{\sigma}_i^2}\right],
\label{eq:likelihood}
\end{equation}
where $\tilde{\sigma}_i^2 = a^2\sigma_{x_i}^2 + \sigma_{y_i}^2 + \sigma_0^2$ is the effective uncertainty of each data point, and the index $i$ runs over all pairs in a given data set. This likelihood is maximized and used to compute confidence intervals for the slope parameter $a$, with the dispersion $\sigma_0$ treated as a nuisance parameter.
The dashed black lines in Fig.~\ref{fig:money_plot} represent the function $f(x)=ax$, where $a$ is the maximum likelihood value of the corresponding data set, and the blue bands correspond to the 1$\sigma$ (68\%) confidence interval around this slope. For our primary (secondary) data set, we find that the maximum-likelihood excess dispersion is $\sigma_0=0.87\pm 0.09$ ($0.9\pm 0.3$) km/sec.

\section{Excess Dispersion of the Data}

Radial orbital motions of binary systems should contribute a dispersion to $\Delta$RV measurements and could be the source of $\sigma_0$. To check this assumption, we have performed a Monte Carlo (MC) analysis by simulating $5\cdot10^4$ binary systems with random orientations, including the effects of $\Delta$RV measurement errors in {\em Gaia}. The mass distribution of stars within the simulation and $\Delta$RV measurement errors are taken from the data. The distribution of semi-major axes is estimated by performing a de-convolution~\citep{1974AJ.....79..745L} of the projected 2D separation of stars, $d$, as measured in the data, under the approximation of circular orbits for all binaries. The distribution of eccentricities is modeled either as a power law~\citep{2017ApJS..230...15M}, such that the distribution scales as $f_e \propto e^{\eta}$ with $\eta = 0$ or $1$, or taken to be zero for all binaries (circular orbits).

The left panel of Fig.~\ref{fig:delta_rv_hist} shows the dispersion of measured $\Delta$RV as a function of $d$ for the primary data set, compared with the MC results. The MC curves are dominated by binary motion at small values of $d$ and by $\Delta$RV measurement errors at large $d$. The data is a good fit to the MC curve, except at intermediate values of $d\approx10^{-2}-10^{-1}$ pc, hinting that binary motion is a significant source of excess dispersion at small separations. However, since the MC underestimates the dispersion at intermediate separations, there is likely an unidentified additional contribution.

One possibility is that the RV errors quoted in the {\em Gaia} catalog are underestimated. The excess dispersion found in this study can be explained if there are experimental uncertainties of $\mathcal{O}(1\ \mathrm{km/sec})$ present in \textit{Gaia} RVS measurements.\\\\

\section{Discussion}

\begin{table}
\centering
\begin{tabular}{cc}
\toprule
systematic checks & slope\\\midrule
primary & $1.26\pm 0.23$ \\
narrow main sequence & $1.14\pm 0.30$ \\ 
observational main sequence & $1.26 \pm 0.26$ \\
$R_{\rm giant}>10~\rsun$ & $1.12 \pm 0.26$ \\
$M_{\rm giant} = 2~\msun$ & $1.34\pm 0.25$ \\
$M_{\rm giant} = 0~\msun$ &  $1.26 \pm 0.35$ \\
$|\Delta\mathrm{RV}|< 3$~km/sec & $1.31 \pm 0.19$\\ 
$|\Delta\mathrm{RV}|< 10$~km/sec & $1.31 \pm 0.25$\\ 
$d > 0$ & $1.27 \pm 0.24$ \\
$d > 10^{-2.5}~\mathrm{pc}$ & $1.48 \pm 0.24$ \\
$\sqrt{\sigma^2_{\mathrm{RV},1} + \sigma^2_{\mathrm{RV},2}} < \infty$ & $1.34 \pm 0.19$ \\
\bottomrule
\end{tabular}
\caption{
Table indicating the systematic variations to our primary data selection and modelling (\textbf{left}) and the resulting slope of the best-fit linear relation (\textbf{right}). The ``narrow main sequence'' variation corresponds to making the main sequence selection bands in Fig.~\ref{fig:HRD} three times narrower. The ``observational main sequence'' variation corresponds to selecting main sequence stars within $\pm0.5$~mag of the fiducial main sequence line identified in \cite{Babusiaux2018}. The ``$R_{\rm giant}>10~\rsun$'' variation corresponds to adjusting the radius selection cut for giant stars to be $R>10~\rsun$. The ``$M_{\rm giant}$'' variations correspond to varying the fixed value used for the masses of all giant stars as indicated. The remaining variations are in the maximum difference in radial velocity, $|\Delta \rm{RV}|$, the minimum distance between the transverse separation between the stars, $d$, and the maximum error in the radial velocity measurement.
\label{tab:checks}} 
\end{table}

The main results of this study are summarized in Fig.~\ref{fig:money_plot}. We find a significant correlation between $\Delta$RV and $\Delta v_{\rm GR} + \Delta v_{\rm CB}$, corresponding to a linear fit of the form $f(x)=ax$, with a slope of order unity. Specifically, for both data sets considered in this study, the best fit slope is consistent with $a=1$ to within measurement errors and inconsistent with zero at the $\sim 5 \sigma$ ($\sim 1.5 \sigma$) level for the primary  (secondary) data set. In the secondary data set, even though the data is more precise, there is a large error in $a$ due to low statistics.

In addition to the results shown in Fig.~\ref{fig:money_plot}, we have performed our analysis with slight variations to the data selection and parameter estimation described above. Details of these variations and the resulting values for $a$ are summarized in Table~\ref{tab:checks}. 
We find that these modifications do not qualitatively change our results.

We find an excess dispersion in the data of order $\sim 0.9$ km/sec that is likely dominated by binary motion at small separations, with some unknown contribution at intermediate separations. A possible explanation is that RV errors in {\em Gaia} are underestimated, however we are unable to pinpoint the origin of the excess dispersion with a high level of certainty.

Moreover, we note that~\textit{Gaia} RVS measurements may exhibit a systematic bias toward high RVs for dim stars~\citep{2019A&A...622A.205K} with corrections up to $0.5~\mathrm{km/sec}$ for apparent magnitudes of $G \approx 12$, and negligible bias for stars brighter than $G \approx 4$. Our primary (secondary) data set exhibits stars with a mean value of $G = 10.15 \, (10.72)$ and standard deviation of $1.50 \, (1.10)$. Since apparent magnitude is correlated with stellar SG, such a systematic bias would serve to enhance, or possibly mimic a GR signal, and might explain the fact that we consistently find $a > 1$ in the analysis. Correcting for this bias, if it exists, is challenging due to the large uncertainty in the size of the effect. We do not attempt a detailed study, but note that taking the bias to vanish for $G<4$ and linearly increasing for $G>4$, with a slope of $(0.5\ \mathrm{km/sec})/(8\ \mathrm{mag})$, reduces the slope in the left panel of  Fig.~\ref{fig:money_plot} to $a = 0.68\pm 0.23$.
One potential discriminator between such a systematic error and a GR signal is that a dependence on $G$ need not manifest as a linear relation in Fig.~\ref{fig:money_plot} (due primarily to the non-trivial relation between apparent magnitude and SG). With future data, one could robustly test if this effect is substantial. 

Systematic uncertainties in the determination of the stellar parameters could also have an impact on our results. For example, a systematic shift in the derived effective temperatures due to unknown reddening would result in different CB corrections than those discussed in Sec.~\ref{sec:cb}. Additionally, CB predictions rely on 3D simulations which are uncertain. Calibration of these models using solar observations revealed discrepancies up to 0.2 km/sec between predictions and observations that may be due to NLTE effects \citep{Gonzalez:2020}. However, it is difficult to estimate this source of uncertainty in the full regime of stellar parameters discussed in this work.

In principle, modifications of the technique developed in this study could be used to test relations such as Eq.~\eqref{eq:M_L_relation} and the $v_{\rm CB}$ fitting formula provided in \cite{2013A&A...550A.103A}. Ultimately, this would be a novel probe of stellar structure and could, for example, assist in an improved understanding of systematics currently prohibiting better measurements of exoplanets and their properties. With the current data, low statistics limit a meaningful test of such models; however, the next {\em Gaia} data release (DR3) is expected to include many more stars with RV measurements and may allow for such an analysis to be performed. In the current study we have shown that a significant signal can be extracted from {\em Gaia} data.

\section{Data Availability}

The data underlying this article are available in the article and in its online supplementary material.

\section*{Acknowledgments}
We thank David Hogg for advice and guidance in the early stages of this study. We also thank David Spergel, Jason Hunt, and George Seabroke for useful conversations. JD was supported in part by the DOE under contract DE-AC02-05CH11231 and in part by the NSF CAREER grant PHY-1915852. OS was supported by the DOE under Award Number DE-SC0007968 and the Binational Science Foundation (grant No. 2018140).

\bibliographystyle{V63/aasjournal}
\bibliography{Comoving_Stars_Bib}

\end{document}